\RequirePackage{fix-cm}
\documentclass[twocolumn]{svjour3}          
\smartqed  

\usepackage[dvipdfmx]{graphicx} 
\usepackage{natbib}
\usepackage{color}
\usepackage{amsmath}
\usepackage{textcomp,gensymb}
\usepackage{comment}
\usepackage{ulem}
\usepackage{upgreek}
\usepackage{gensymb}
\usepackage{hyperref}
\hypersetup{
    colorlinks=true,
    linkcolor=blue,
    filecolor=blue,      
    urlcolor=blue,
    }

\usepackage{ulem}

 \journalname{}

\begin{document}
\title{Flow birefringence of cellulose nanocrystal suspensions in three-dimensional flow fields: revisiting the stress-optic law}

\subtitle{}

\author{Kento Nakamine$^{1}$, Yuto Yokoyama$^{1}$, William Kai Alexander Worby$^{2}$,  \\Masakazu Muto$^{3}$, Yoshiyuki Tagawa$^{1*}$}

\authorrunning{Kento Nakamine, Yuto Yokoyama, William Kai Alexander Worby, Masakazu Muto, Yoshiyuki Tagawa}

\institute{
1: Department of Mechanical Systems Engineering, Tokyo University of Agriculture and Technology, Koganei, Tokyo, 184-8588, Japan\\
2: Department of Industrial Technology and Innovation, Tokyo University of Agriculture and Technology, Koganei, Tokyo, 184-8588, Japan\\
3: Department of Electrical and Mechanical Engineering, Nagoya Institute of Technology, Nagoya, Aichi, 466-8555, Japan\\
$^{*}$Corresponding author. \email{tagawayo@cc.tuat.ac.jp}\\
}
\date{Received: date / Accepted: date}

\maketitle


\begin{abstract}
This study systematically investigates the flow birefringence of cellulose nanocrystal (CNC) suspensions.
The aim is to clarify the importance of the stress component along the camera's optical axis in the stress-optic law (SOL), which describes the relationship between birefringence, the retardation of transmitted polarized light, and the stress field.
More than 100 datasets pertaining to the retardation of CNC suspensions (concentrations of 0.1, 0.3, 0.5, and 1.0 wt\%) in a laminar flow field within a rectangular channel (aspect ratios of 0.1, 1, and 3) are systematically obtained.
The measured retardation data are compared with the predictions given by the conventional SOL excluding the stress component along the camera's optical axis and by the SOL including these components as second-order terms (2nd-order SOL).
The results show that the 2nd-order SOL gives a significantly better agreement with the measurements.
Based on the 2nd-order SOL, the retardation at the center of the channel, where the effect of the stress component along the camera's optical axis is most pronounced, is predicted to be proportional to the square of the flow rate, which agrees with the experimental data.
The results confirm the importance of considering the stress component along the camera's optical axis in the flow birefringence of CNC suspensions at high flow rates, even for quasi-two-dimensional channel flow.

\end{abstract}

\section{Introduction}\label{sec1}

Cellulose nanocrystals (CNCs) are rod-shaped crystals with a high aspect ratio \citep{Beck-Candanedo2005}.
These rigid, highly crystalline nanoparticles are made from wood and plants~\citep{Habibi2010},
and have characteristics including a low density and high strength \citep{Saito2013}, as well as interesting thermodynamic and rheological properties~\citep{Buffa2019}.
Many researchers have discussed the potential applications of such properties~\citep{Thiruganasambanthan2022,Mu2019,Lagerwall2014}.

One property of CNCs that we are interested in is birefringence \citep{kadar2021,detert2023}.
When a suspension containing particles with a large aspect ratio, such as CNCs, is at rest, the particles within the fluid are randomly oriented, resulting in optical isotropy.
As the fluid flows, the shear forces orient the particles in a preferred direction, resulting in the fluid exhibiting birefringence \citep{aben1997,kadar2021}.
This birefringence property of a fluid (or flow birefringence) appears when the particles are under a mechanical force~(corresponding to shear in our case) and are aligned in a uniform direction~\citep{hausmann2018, calabrese2021structure}.
Flow birefringence strongly and directly depends on the stress applied to the bulk fluid; details are given in Sec.~\ref{subsec2}.
Therefore, quantitative measurements of flow birefringence provide a direct estimation of the fluid stress field, without the need for velocity field measurements through techniques such as particle image velocimetry.

If light is passed through a fluid while it is subject to shear-induced birefringence, the light is elliptically polarized with a retardation $\Delta$ that depends on the magnitude of the birefringence and an orientation $\phi$ that depends on the direction of the particle.
The relationship between $\Delta$, $\phi$, and the stresses in the material is called the stress-optic law (SOL) \citep{aben1993a,ramesh2021}.
Using the SOL, the stress can be estimated from the retardation $\Delta$ and orientation $\phi$ of the emitted polarized light obtained by optical measurements \citep{prabhakaran1975stress,janeschitz2012polymer,knight1976orthotropic,noto2020applicability}.
Studies using CNC suspensions have reported the applicability of the SOL in quasi-two-dimensional flow fields, such as rectangular channels, in which the channel depth is greater than the width, or in Taylor--Couette flows \citep{lane2022,calabrese2021effects,calabrese2022,lane2023}, as a means of understanding the flow birefringence of such fluids.
Studies on flow birefringence in three-dimensional flow fields are limited compared with those on two-dimensional flows.

In the three-dimensional case, previous studies have suggested that the SOL exhibits a nonlinear relationship between the retardation and stresses \citep{McAfee1974,doyle1982nonlinearity,aben1997}.
\citet{McAfee1974} observed a three-dimensional flow in a channel and stated that ``the isoclinic [retardation in this paper] was dependent on shear strain rates both normal and parallel [camera's optical axis in this paper] to the light ray'', but they did not derive the direct relation between the SOL and their observations. 
\citet{doyle1982nonlinearity} extended the SOL to three-dimensional flows by including the stress component along the camera's optical axis as second-order terms.
This study highlighted the importance of the stress component along the camera's optical axis, although the theory was not validated against experiments.
The effect of the stress component along the camera's optical axis in the SOL has been neglected in the birefringence of solids, i.e., photoelasticity.
\citet{aben1997} discussed the flow birefringence by referring to Doyle's work, although their extended SOL neglected the stress component along the camera's optical axis.
Other significant works \citep{ainola1998application,kim2017monitoring,ober2011spatially,clemeur2004} have also omitted the stress component along the camera's optical axis.
Thus, the influence of the stress component along the camera's optical axis on flow birefringence has not been incorporated into many previous studies, although its importance has been identified.
The negative effects of this omission seem to have been avoided by using a quasi-two-dimensional flow field, as described above.
Nevertheless, the results of several studies using quasi-two-dimensional channels highlight the discrepancies between the retardation predicted by the SOL for the two-dimensional case and the experimental data, especially in the center of the channel \citep{kim2017monitoring,ober2011spatially,calabrese2023a}.
Thus, to clarify the effect of the stress component along the camera's optical axis on flow birefringence, it is necessary to revisit the extended SOL described by \citet{doyle1982nonlinearity} and verify it experimentally.
To the best of the authors' knowledge, the SOL considering the stress component along the camera's optical axis has never been experimentally validated, although fluid flows are always strictly three-dimensional.

\begin{figure}[t]
\centering
\includegraphics[width=1\columnwidth]{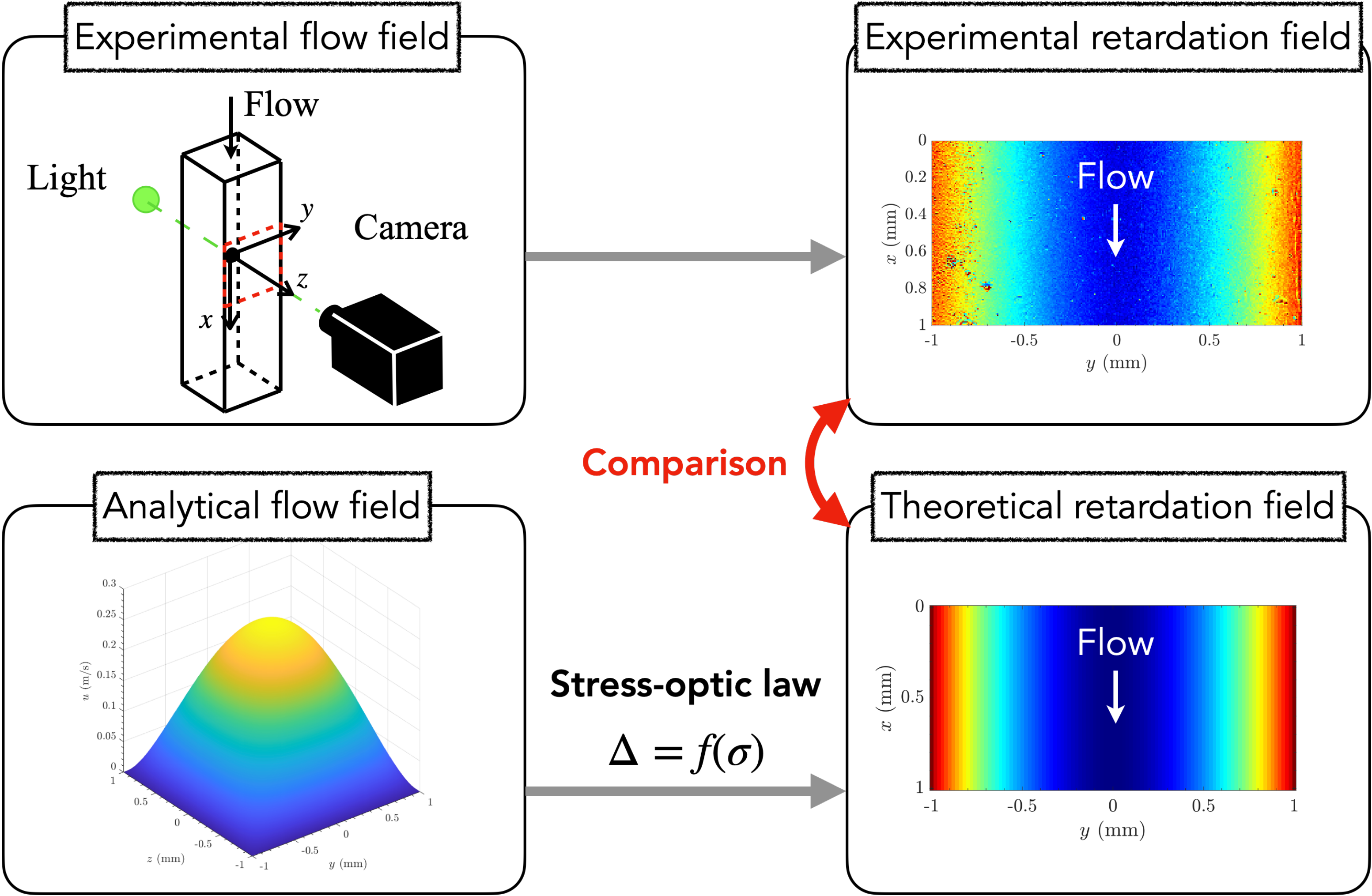}
\caption{Schematic of the concept of this study. The theoretical retardation field, calculated using an SOL from the analytical solution of the flow field, is compared with the experimentally obtained retardation field to verify the SOL.}
\label{fig:Concept}
\end{figure}

This study determines the effect of the stress component along the camera's optical axis on the flow birefringence of CNC suspensions by revisiting Doyle's SOL \citep{doyle1982nonlinearity}.
For this purpose, we performed systematic experiments measuring the flow birefringence of CNC suspensions in steady laminar flow in rectangular channels.
The retardation field measured during the experiments was compared with that predicted by the SOL proposed by \citet{doyle1982nonlinearity}.
The methodology used in this study is shown in Fig. \ref{fig:Concept}.
A steady laminar flow of a Newtonian fluid in a rectangular channel is considered because analytical solutions for the stress field are available for such cases.
The aspect ratio of the rectangular channel was varied to change the stress component's contribution along the camera's optical axis.

\section{Methodology: Polarization measurements in rectangular channel flow of CNC suspensions}\label{sec2}

In this section, we explain the SOL, which is the relationship between retardation and stress component.
We then explain how retardation is measured and how the theoretical retardation of a fluid is calculated.

\subsection{Stress-optic law}\label{subsec2}

In this section, we carefully review the SOLs described by Aben and Doyle \citep{aben1997,doyle1982nonlinearity}.
We consider which formulations include three-dimensional effects, i.e., the effect of the stress component along the camera's optical axis, and refer to 1st-order SOLs for those that do not include the stress component along the camera's optical axis and 2nd-order SOLs for those that do include the stress component along the camera's optical axis.

The retardation of light causes birefringence due to the anisotropic refractive index tensor of a birefringent material when loaded by stress.
The material's refractive index is determined by the relative permittivity, also known as the dielectric constant.
If $z$ is set as the camera's optical axis in Cartesian coordinates, the dielectric tensor $\boldsymbol{\varepsilon}$ can be written as
\begin{equation}
\boldsymbol{\varepsilon} = \begin{bmatrix}
\varepsilon_{xx} & \varepsilon_{xy} & \varepsilon_{xz} \\
\varepsilon_{yx} & \varepsilon_{yy} & \varepsilon_{yz} \\
\varepsilon_{zx} & \varepsilon_{zy} & \varepsilon_{zz}
\end{bmatrix}.
\end{equation}
The retardation $\Delta$ and orientation $\phi$ of the polarized light through the material are related to the components of the dielectric tensor in Cartesian coordinates as follows \citep{aben1997,doyle1982nonlinearity}:
\begin{eqnarray}\label{eq:delta-dielectric}
V_1 \equiv \Delta \cos 2 \phi &=& \frac{1}{2n_0} \int \left(\varepsilon_{yy}-\varepsilon_{xx}\right) dz, \label{eq:delta-dielectric1}\\
V_2 \equiv \Delta \sin 2 \phi &=& \frac{1}{n_0} \int \varepsilon_{xy} dz,\label{eq:delta-dielectric2}
\end{eqnarray}
where $n_0$ is the initial refractive index of the material.
Therefore, $\Delta$ and $\phi$ can be expressed as
\begin{eqnarray} \label{eq:photoelastic_parameters}
\Delta = \sqrt{V_1^2 + V_2^2}, \qquad \phi = \frac{1}{2} \tan^{-1} \frac{V_2}{V_1}.
\end{eqnarray}
If $\Delta$ is less than $\lambda/4$ and the rotation of the principal stress direction does not exceed $\pi/6$, Eqs. (\ref{eq:delta-dielectric1}) and (\ref{eq:delta-dielectric2}) are valid \citep{aben1989,aben1993a}.
This approximation holds for the conditions considered in this study because the measured retardation is always less than $\lambda/4$.
The optical effect is a function of the strain rate $\dot e_{jk}$ \citep{aben1997,doyle1982nonlinearity}:
\begin{equation}
\frac{1}{2n_0} \varepsilon_{jk} = f(\dot e_{jk}).
\end{equation}
Using the Cayley--Hamilton theorem \citep{aben1997,doyle1982nonlinearity}, the optical effect can be written as
\begin{equation}
\frac{1}{2n_0} \varepsilon_{jk} = \alpha_0 \delta_{jk} + \alpha_1 \dot e_{jk} + \alpha_2 \dot e_{jl} \dot e_{lk},
\end{equation}
where $\alpha_{0}$, $\alpha_{1}$, and $\alpha_{2}$ are material-specific constants and $\delta_{jk}$ is the Kronecker delta.
Here, $j$, $k$, and $l$ are Einstein's notation indices.
As described in previous studies \citep{aben1997, doyle1982nonlinearity}, the relationship between the retardation, orientation, and the components of the strain rate tensor can be expressed using Eqs. (\ref{eq:delta-dielectric1}) and (\ref{eq:delta-dielectric2}) as:
\begin{equation}
\begin{split}
V_1
&=
\Delta\cos2\phi
=
\int \left\{\alpha_1\left(\dot e_{yy} - \dot e_{xx}\right) \right. \\
&\left.+ \alpha_2\left[\left(\dot e_{yy} + \dot e_{xx}\right)\left(\dot e_{xx} - \dot e_{yy}\right)
+\dot e_{yz}^{2} - \dot e_{xz}^{2}
\right]\right\}dz ,
\end{split}
\end{equation}
\begin{equation}
\begin{split}
V_2
=
\Delta\sin2\phi
&=
\int \left\{2\alpha_1\dot e_{xy} \right. \\
&\left. +\alpha_2\left[
2\left(\dot e_{yy} + \dot e_{xx}\right)\dot e_{xy}+2\dot e_{yz}\dot e_{xz}\right]\right\}dz.
\end{split}
\end{equation}

For a Newtonian fluid, the viscous stress is proportional to the strain rate, i.e., $\sigma_{jk} = \eta \dot e_{jk}$, where $\eta$ is the liquid viscosity.
Thus, these equations become
\begin{equation}
\begin{split}
V_1
&=
\Delta\cos2\phi
=
\int \left\{ C_1\left(\sigma_{yy} - \sigma_{xx}\right) \right. \\
& \left.+ C_2\left[\left(\sigma_{yy} + \sigma_{xx}\right)\left(\sigma_{yy} - \sigma_{xx}\right)
+\sigma_{yz}^{2} - \sigma_{xz}^{2}
\right] \right\} dz \label{eq:v1},
\end{split}
\end{equation}
\begin{equation}
\begin{split}
V_2
=
\Delta\sin2\phi
&=
\int \left\{ 2C_1\sigma_{xy} \right. \\
&\left.+C_2\left[
2\left(\sigma_{yy} + \sigma_{xx}\right)\sigma_{xy}+2\sigma_{yz}\sigma_{xz}
\right] \right\} dz. \label{eq:v2}
\end{split}
\end{equation}
Note that $C_1 = \alpha_1/\eta$ and $C_2 = \alpha_2/\eta^2$.

As described in Section \ref{sec1}, the second-order terms of the stress component along the camera's optical axis in Eqs. (\ref{eq:v1}) and (\ref{eq:v2}), i.e., $\sigma_{xz}$ and $\sigma_{yz}$, have been neglected in previous studies of flow birefringence, which assumed that $C_2 = 0$ \citep{aben1997,ober2011spatially}.
This approach leads to the following well-known equations of integrated photoelasticity for solids \citep{aben1993a,yokoyama2023a}:
\begin{eqnarray}
V_1
&=&
\Delta\cos2\phi
=C_1 \int \left(\sigma_{yy} - \sigma_{xx}\right) dz, \label{eq:local retardation1}\\
V_2
&=&
\Delta\sin2\phi
=2 C_1 \int \sigma_{xy} dz. \label{eq:local retardation2}
\end{eqnarray}
Hereafter, we refer to Eqs. (\ref{eq:local retardation1}) and (\ref{eq:local retardation2}) as the ``1st-order SOL'', often simply called the SOL in conventional studies.
This formulation excludes the stress component along the camera's optical axis.
In the remainder of this paper, we call Eqs. (\ref{eq:v1}) and (\ref{eq:v2}), in which $C_2 \neq 0$, the ``2nd-order SOL''. This formulation includes the stress component along the camera's optical axis.

Based on 1st- and 2nd-order SOLs, we can calculate the retardation $\Delta$ if the stress field is known a priori.

\subsection{Working fluids}

\begin{figure}[t]
\centering
{\includegraphics[width=1\columnwidth]{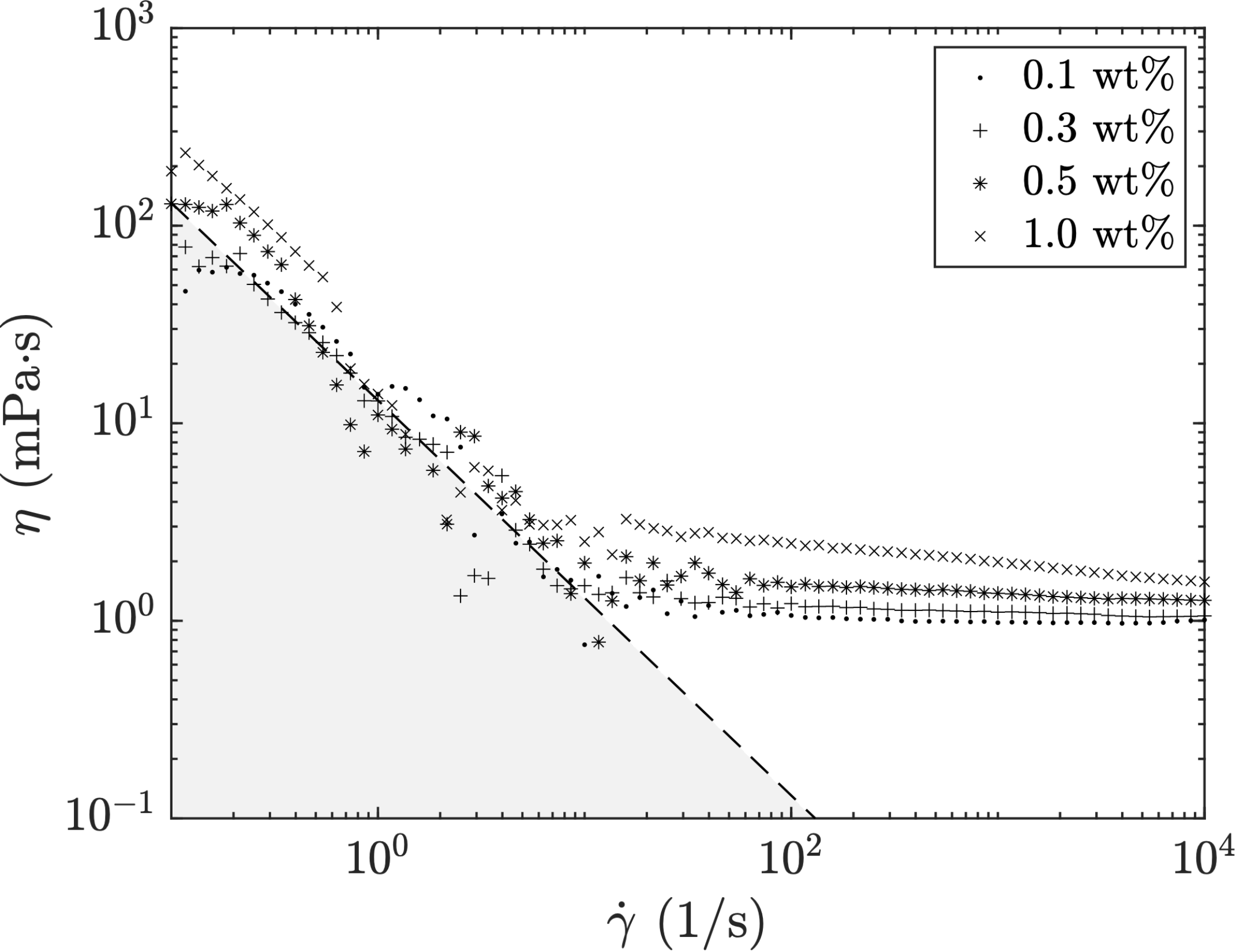}}
\caption{\label{viscosity}Shear viscosity $\eta$ versus shear rate $\dot{\gamma}$ of CNC suspensions with concentrations $c$ of 0.1, 0.3, 0.5, and 1.0 wt\% and ultra-pure water, as measured by a rheometer. The dashed line indicates the low-torque limit of the rheometer.}

\end{figure}

The working fluids disperse CNCs (CNC-HSFD, Cellulose Lab, Ltd.) in ultra-pure water.
The CNCs are mixed with ultra-pure water using a magnetic stirrer (CHPS-170DF, ASONE Co., Ltd.) at 25 $^\circ$C and 600 rpm for at least 1 h.
The CNC suspensions are then sonicated using an ultrasonic processor (UX-300, Mitsui Electric Co. Ltd.) for 10 min, which reduces the error between the experimental and theoretical orientation \citep{lane2022,ober2011spatially}.
CNC suspensions with concentrations of 0.1, 0.3, 0.5, and 1.0 wt\% are used in the experiments.
Figure \ref{viscosity} shows the shear viscosity $\eta$ versus the shear rate $\dot{\gamma}$ of the CNC suspensions and ultra-pure water, as measured using a rheometer (MCR302, Anton Paar Co., Ltd.).
The shear viscosities at $\dot{\gamma}=10^3$ 1/s are 1.01, 1.06, 1.27, and 1.58 mPa$\cdot$s for 0.1, 0.3, 0.5, and 1.0 wt\% CNC suspensions, respectively.
These values are used when the theoretical flow field is calculated.
There is negligible interparticle interaction in the CNC suspensions if the concentration is less than 0.5 wt\% \citep{lane2022,bertsch2019}.
The shear rheology of the 1.0 wt\% CNC suspension exhibits a slight shear-thinning viscosity.
We evaluated the effect of this weak shear-thinning property on the flow field through numerical simulations using commercial software (COMSOL Multiphysics).
As described in detail in \ref{sec:non-Newtonian effect}, its flow field is very similar to that of a Newtonian fluid.
Therefore, all CNC suspensions used in this study can be considered as Newtonian fluids.

\subsection{Experiments}

\begin{figure}[t]
\centering
\includegraphics[width=1\columnwidth]{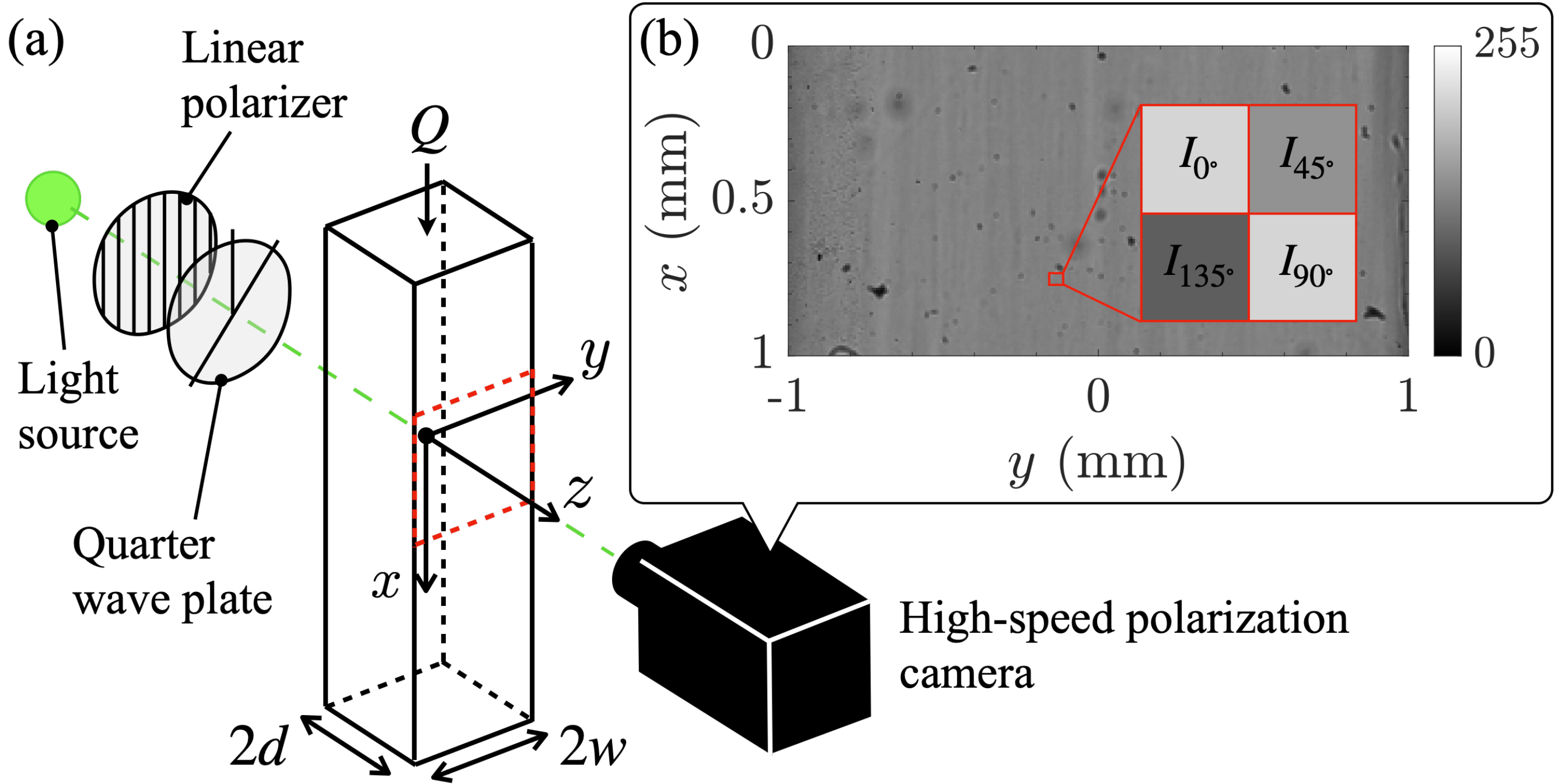}
\caption{\label{experimental setup}(a) Schematic diagram of the experimental setup. (b) Raw image taken by the polarization camera. The inset shows that the neighboring pixels are polarizers with different directions. The retardation is calculated from the intensity values of the four pixels next to each other.}
\end{figure}

A schematic diagram of the experimental setup is shown in Fig. \ref{experimental setup}, along with the coordinates of the system.
A linear polarizer, a quarter-wave plate, and a rectangular channel are placed between a light source (SOLIS-565C, Thorlabs), which generates green light with a typical wavelength $\lambda$ of 540 nm, and a high-speed polarization camera (CRYSTA PI-5WP, Photron, temporal resolution: 250 fps, spatial resolution: 848 $\times$ 680 pixels for an 8-bit light intensity image).
The light passes through the linear polarizer at $0^\circ$ and the quarter-wave plate at $45^\circ$ with respect to the $y$-axis, and becomes circularly polarized.
As the light passes through the stress field, it accumulates a retardation and orientation corresponding to the stress states encountered along the camera's optical axis.
After passing through the stress field, the light becomes elliptically polarized and is recorded by the polarization camera, which contains ``super-pixels'' \citep{lane2022a}.
Each super-pixel of the image sensor within the polarization camera consists of four linear polarizers with four different directions: $0^\circ$, $45^\circ$, $90^\circ$, and $135^\circ$.
The light intensity values measured by the camera's sensor through each linear polarizer are denoted as $I_{0^\circ}, I_{45^\circ}, I_{90^\circ}$, and $I_{135^\circ}$, respectively, as shown in Fig. \ref{experimental setup}(b).

\begin{figure}[t]
\centering
\includegraphics[width=1\columnwidth]{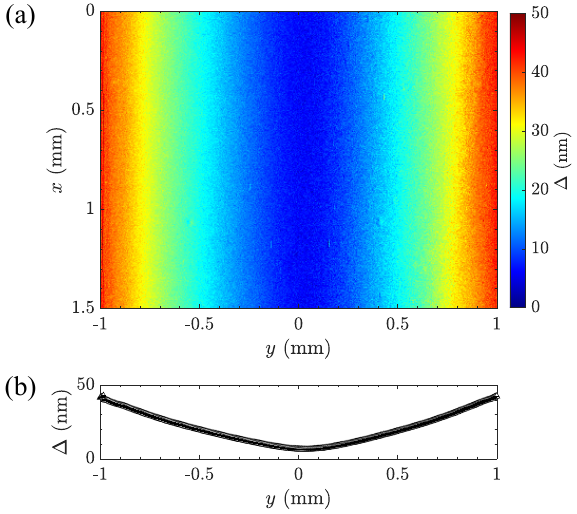}
\caption{\label{Fig:MeasurementData} (a) Spatial distribution of the measured retardation field $\Delta$ of CNC 0.5 wt\% in the case of $Q=30$ ml/min with an aspect ratio of $w/d=1$. (b) Line profile of the spatiotemporally averaged retardation of (a) over a period of 0.4 s.}
\end{figure}

\begin{figure}[t]
\centering
\includegraphics[width=1\columnwidth]{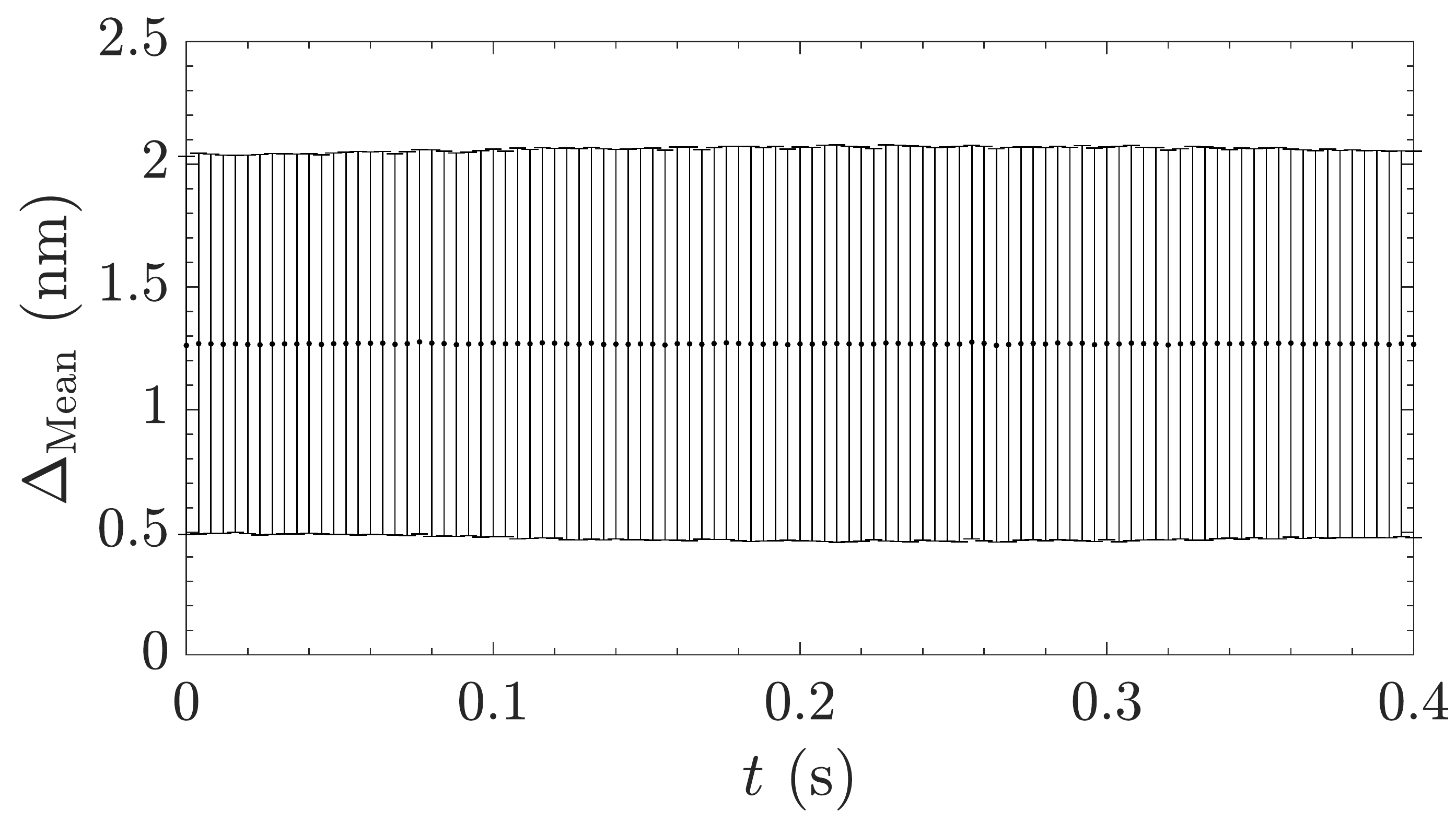}
\caption{\label{Fig:retardation_at_rest} Temporal evolution of spatially averaged retardation at rest for 0.5 wt\% CNC suspension and aspect ratio of 1. Error bars indicate the standard deviation of the calculated average retardation over the entire image.}
\end{figure}

Using the four-step phase-shifting method \citep{ramesh2021,otani1994,onuma2014,yokoyama2023,miyazaki2021}, the retardation $\Delta$ is obtained from the four intensity values of the super-pixels as follows:
\begin{gather}
\Delta = \frac{\lambda}{2\pi} \sin^{-1}{\frac{\sqrt{\left(I_{90^\circ}-I_{0^\circ}\right)^2+\left(I_{45^\circ}-I_{135^\circ}\right)^2}}{I/2}}, \label{eq:delta}
\end{gather}
where $I = I_{0^\circ} + I_{45^\circ} + I_{90^\circ} + I_{135^\circ}$.
An example of the measured retardation calculated using software (Photron Ltd., CRYSTA Stress Viewer) is shown in Fig. \ref{Fig:MeasurementData}(a).
The spatial resolution of the retardation data is 424 $\times$ 340 pixels, which is a quarter of 848 $\times$ 680 pixels.
The background noise is subtracted before measuring the retardation field during fluid flow.
To reduce the shot noise of the camera, the measured retardation is spatially averaged in the flow direction ($x$-direction) and temporally averaged over $0.4$ s, resulting in a line profile (Fig. \ref{Fig:MeasurementData}(b)).
As discussed later, the flow field is effectively constant in the flow direction and time.
We verified the lowest measurable retardation of the polarization camera used in this study
by measuring the retardation field of the rectangular channel filled by a non-flowing CNC suspension.
Figure \ref{Fig:retardation_at_rest} shows the temporal evolution of the spatially averaged retardation $\Delta_{\rm Mean}$, which exhibits retardation of approximately $1.27$ nm.
This is considered to be the lowest measurable retardation of the polarization camera.

We prepared channels with three aspect ratios $w/d$, i.e., $w/d = 0.1, 1$, and $3$.
The aspect ratio $w/d = 0.1$ was produced using a channel width of $2w = 1$ mm and a channel depth of $2d = 10$ mm.
The aspect ratio $w/d = 1$ was produced using a channel width of $2w = 2$ mm and a channel depth of $2d = 2$ mm.
The aspect ratio $w/d = 3$ was produced using a channel width of $2w = 6$ mm and a channel depth of $2d = 2$ mm.
The rectangular channel with $w/d = 1$ was made of quartz glass.
The rectangular channels with aspect ratios of $w/d = 0.1$ and $3$ were prepared using a three-dimensional printer (Form3, Formlabs; material: clear resin).
To create a flat surface, the front and back sides of the channels were polished.
The flow channels were sealed with two glass plates level with the surfaces.
The flow of the channel with $w/d = 0.1$ can be considered quasi-two-dimensional, as mentioned above.

The coordinate system is set up as shown in Fig. \ref{experimental setup}. 
The length of the channel is around 50 mm.
The fluid was supplied using a syringe pump (Pump 11 Elite, Harvard Apparatus) to produce a steady laminar flow.
The flow rates in the experiments ranged from 1--100 ml/min.
The Reynolds number $Re = \rho U_{mean} R_H / \eta$ reaches a maximum value of approximately 200 when the flow rate is 100 ml/min using the channel with $w/d = 1$.
Here, $U_{mean}$ is the mean velocity, defined by $U_{mean} = Q/(4wd)$, and $R_H$ is the hydraulic radius, defined by $R_H = wd/(w+d)$.
The measurement area extends over $x \sim 45$ mm.
The entrance length $L_e$, which is defined as $L_e = 0.06 Re R_H$, is sufficiently smaller than the distance between the flow inlet and the measurement position in all experimental conditions.
Therefore, the flow field is effectively fully developed.

\subsection{Theoretical flow field}

\begin{figure*}[t]
\centering
\includegraphics[width=1\textwidth]{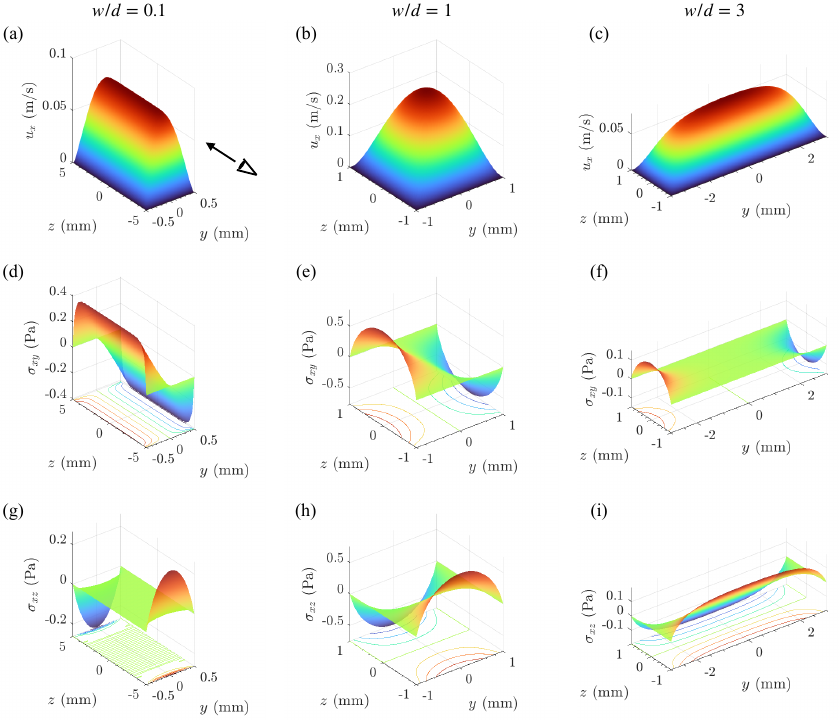}
\caption{\label{fig:u}(a,b,c) Velocity fields for different aspect ratios $w/d$. (d,e,f) Distributions of shear stress $\sigma_{xy}$ for different aspect ratios $w/d$. (g,h,i) Distributions of shear stress $\sigma_{xz}$ for different aspect ratios $w/d$. All quantities calculated using $Q = 30$ ml/min, $\eta = 1.12$ mPa$\cdot$s.}
\end{figure*}

This section describes the theoretical flow field in a rectangular channel to calculate the retardation and orientation field to be compared with experiments.
First, the velocity distribution of a steady laminar flow in a rectangular channel is derived from the Navier--Stokes equations as follows \citep{delplace2018laminar, 2006}:
\begin{gather}
\left( \frac{\partial^2}{\partial y^2}+\frac{\partial^2}{\partial z^2}\right)u_x =-\frac{4 Q}{w d^3 K}, \label{eq:navi} \\
u_y = u_z = 0,
\end{gather}
where
\begin{equation}
K = \frac{16}{3} - \frac{1024}{\pi^5}\frac{d}{w}
\sum_{n=0}^\infty \frac{1}{(2n+1)^5}
\tanh{\frac{\left(2n+1\right)\pi w}{2d}}.
\label{eq:K}
\end{equation}
Here, $Q$ is the flow rate and $w$, $d$ are the channel width and channel depth, respectively.
$u_y$ and $u_z$ are the velocities in the $y$- and $z$-directions, respectively.
The velocity distribution $u_x(y,z)$, which satisfies Eqs. (\ref{eq:navi})--(\ref{eq:K}), and the no-slip boundary condition at the wall ($y = \pm w$ or $z = \pm d$) are defined as follows:
\begin{equation}
\begin{split}
u_x(y,z) &= \frac{64Q}{w d \pi^3 K} \sum_{n=0}^\infty \frac{(-1)^n}{(2n+1)^3} \\
&\left[1-\dfrac{\cosh \dfrac{(2n+1)\pi y}{2d}}{\cosh\dfrac{(2n+1)\pi w}{2d}}\right]\cos\dfrac{(2n+1)\pi z}{2d}, \label{eq:u} 
\end{split}
\end{equation}
\begin{equation}
u_y = u_z = 0. \label{eq:v,w}
\end{equation}
Examples of the theoretical velocity distributions with different aspect ratios $w/d$ are shown in Figs. \ref{fig:u}(a,b,c).
From Eqs. (\ref{eq:u}) and (\ref{eq:v,w}), the viscous stresses are calculated as follows:
\begin{equation}
\begin{split}
\sigma_{xy}&(y,z) = \eta \frac{\partial u_x}{\partial y} = \frac{64Q\eta}{w d \pi^3 K} \sum_{n=0}^\infty \dfrac{(-1)^n}{(2n+1)^3} \\ 
&\dfrac{(2n+1)\pi}{2d} \dfrac{\sinh \dfrac{(2n+1)\pi y}{2d}}{\cosh\dfrac{(2n+1)\pi w}{2d}}\cos\dfrac{(2n+1)\pi z}{2d}, \label{eq:o_xy}
\end{split}
\end{equation}
\begin{equation}
\begin{split}
&\sigma_{xz}(y,z)=\eta \frac{\partial u_x}{\partial z} = \frac{64Q\eta}{w d \pi^3 K} \sum_{n=0}^\infty \dfrac{(-1)^n}{(2n+1)^3} (-1) \\
& \dfrac{(2n+1)\pi}{2d} \left[1-\dfrac{\cosh \dfrac{(2n+1)\pi y}{2d}}{\cosh\dfrac{(2n+1)\pi w}{2d}}\right] \sin\dfrac{(2n+1)\pi z}{2d}, \label{eq:o_xz}
\end{split}
\end{equation}
\begin{equation}
\sigma_{xx} = \sigma_{yy} = \sigma_{zz} = \sigma_{yz} = 0. \label{eq:o_xx}
\end{equation}
Examples of the theoretical stress fields in the channel are shown in Figs. \ref{fig:u}(d,e,f) and \ref{fig:u}(g,h,i).
Therefore, Eqs. (\ref{eq:v1}) and (\ref{eq:v2}) in this flow field become
\begin{gather}
V_1
=
\Delta\cos2\phi
=
- C_2 \int \sigma_{xz}^{2} dz \label{eq:v1_rev}, \\
V_2
=
\Delta\sin2\phi
=
2 C_1 \int \sigma_{xy} dz. \label{eq:v2_rev}
\end{gather}
In particular, the stress component at the wall ($y=\pm w$) and at the channel center ($y=0$) are
\begin{equation}
\begin{split}
\sigma_{xy}&(\pm w,z) = \frac{64Q\eta}{w d \pi^3 K} \sum_{n=0}^\infty \dfrac{(-1)^n}{(2n+1)^3} (\pm 1) \\
&\dfrac{(2n+1)\pi}{2d} \tanh \dfrac{(2n+1)\pi w}{2d}\cos\dfrac{(2n+1)\pi z}{2d}, \label{eq:o_xy_wall}
\end{split}
\end{equation}
\begin{eqnarray}
&&\sigma_{xz}(\pm w,z) = 0, \label{eq:o_xz_wall}\\
&&\sigma_{xy}(0,z) = 0, \label{eq:o_xy_center}
\end{eqnarray}
\begin{equation}
\begin{split}
&\sigma_{xz}(0,z) = \dfrac{64Q\eta}{w d \pi^3 K} \sum_{n=0}^\infty \dfrac{(-1)^n}{(2n+1)^3} (-1) \\
&\dfrac{(2n+1)\pi}{2d} \left[1-\dfrac{1}{\cosh\dfrac{(2n+1)\pi w}{2d}}\right]\sin\dfrac{(2n+1)\pi z}{2d}. \label{eq:o_xz_center} 
\end{split}
\end{equation}
Thus, the theoretical retardation and orientation can be calculated using Eqs. (\ref{eq:photoelastic_parameters}), (\ref{eq:o_xz})--(\ref{eq:v2_rev}) and compared with the experimental data.
In particular, the scaling of the retardation at the wall ($y = \pm w$), $\Delta_{\rm wall}$, and the retardation at the center ($y = 0$), $\Delta_{\rm center}$, can be calculated as follows:
\begin{eqnarray}
\Delta_{\rm wall} = \left| 2C_1 \int \sigma_{xy}(\pm w,z) dz \right| \propto Q, \label{eq:Delta_wall} \\
\Delta_{\rm center} = \left|-C_2 \int \sigma_{xz}^2(0,z) dz\right| \propto Q^2. \label{eq:Delta_center}
\end{eqnarray}
These equations suggest that the retardation at the wall increases in proportion to the 1st power of $Q$ and the retardation at the center increases in proportion to the 2nd power of $Q$.
Furthermore, if we use the 1st-order SOL, i.e., $C_2 = 0$, the retardation is expected to be zero at the center of the channel.

In terms of the effects of stress components along the camera's optical axis on flow birefringence, i.e., $C_2$ term in the SOL, the effect is most pronounced at the center of the channel.
The effect is negligible at the wall.
In other words, if the effect is not ignored, the retardation at the center of the channel will not be zero, although the retardation at the wall may not be affected. 
Therefore, the retardation at the wall and at the center of the channel are the main targets of discussion in this study.

\begin{figure*}[t]
\centering
\includegraphics[width=1\textwidth]{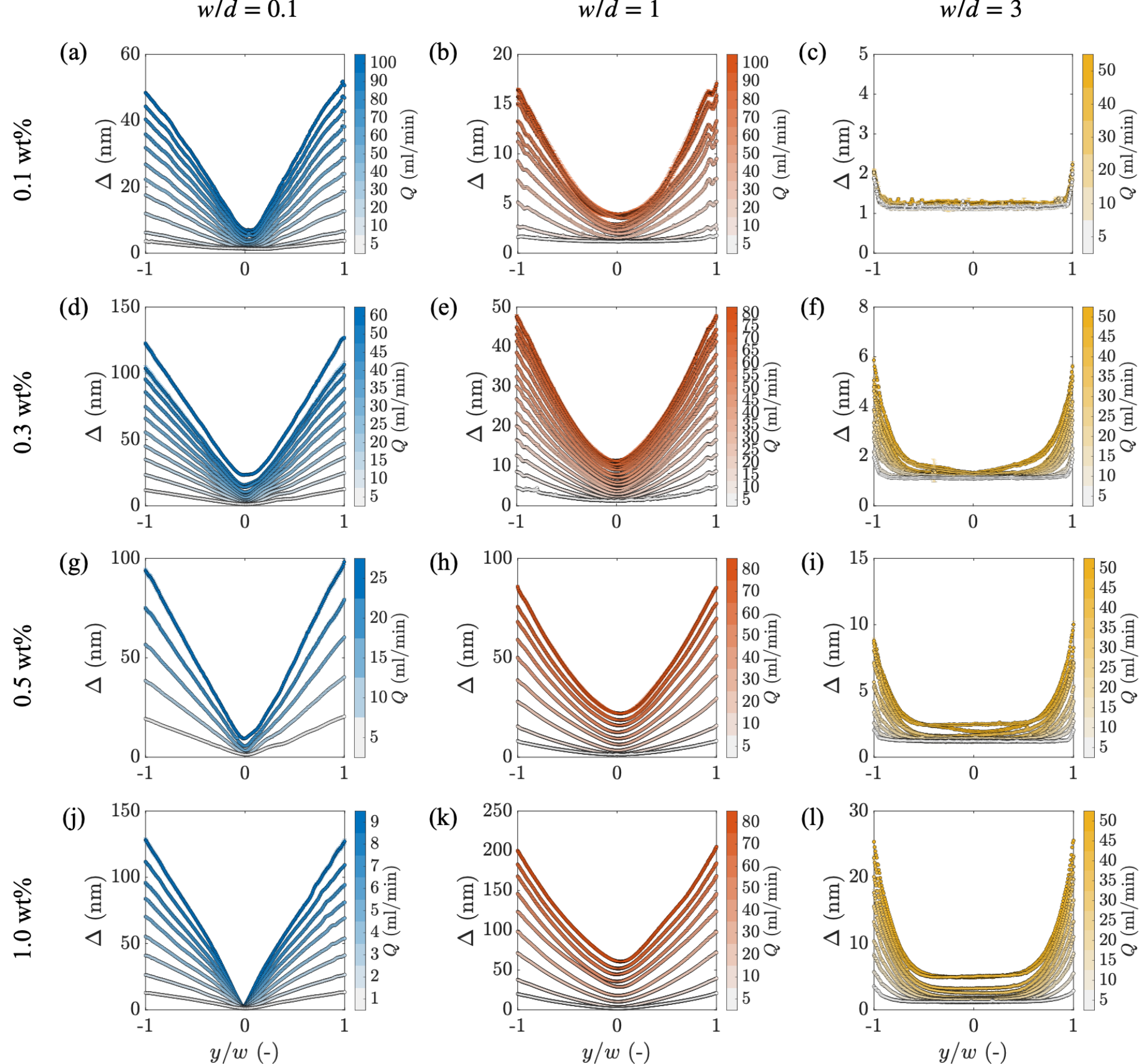}
\caption{\label{fig:retardation_line} Line profiles of the retardation for different aspect ratios, CNC concentrations, and flow rates. The horizontal axis is normalized by the width of the channel $w$. The error bars show the standard deviation calculated when the retardation is temporally averaged over 250 frames, but the size of the marker tends to obscure these quantities.}
\end{figure*}

\section{Results and discussion}\label{sec3}

\begin{figure*}[t]
\centering
\includegraphics[width=0.8\textwidth]{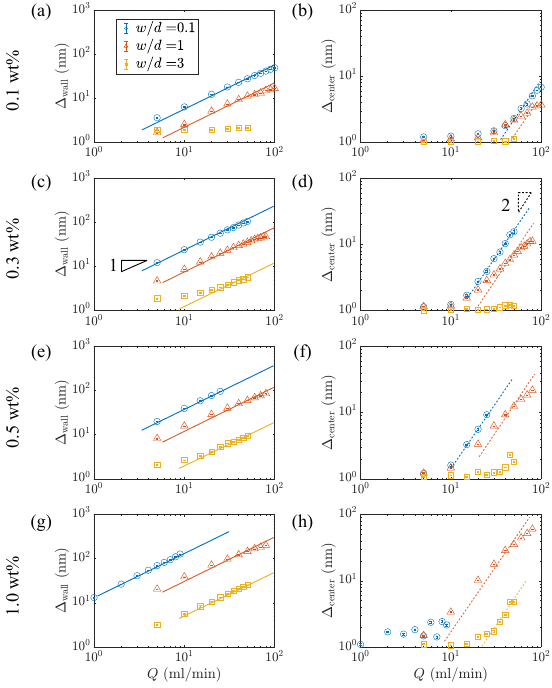}
\caption{\label{fig:retardation_wall_center} (Left column) Relationship between retardation at $y = \pm w$ $\Delta_{\rm wall}$ and flow rate $Q$ with different CNC concentrations: (a) 0.1 wt\%, (c) 0.3 wt\%, (e) 0.5 wt\%, and (g) 0.1 wt\%. (Right column) Relationship between retardation at the center $\Delta_{\rm wall}$ and flow rate $Q$ with different CNC concentrations: (a) 0.1 wt\%, (c) 0.3 wt\%, (e) 0.5 wt\%, and (g) 0.1 wt\%. Solid lines in (a,c,e,g) have a slope of 1, indicating $\Delta_{\rm wall} \propto Q$ predicted by Eq. (\ref{eq:Delta_wall}). Dashed lines in (b,d,f,h) have a slope of 2, indicating $\Delta_{\rm center} \propto Q^2$ predicted by Eq. (\ref{eq:Delta_center}). The error bars show the standard deviation calculated when the measured retardation is temporally averaged over 250 frames.}
\end{figure*}

\begin{figure*}[t]
\centering
\includegraphics[width=0.8\textwidth]{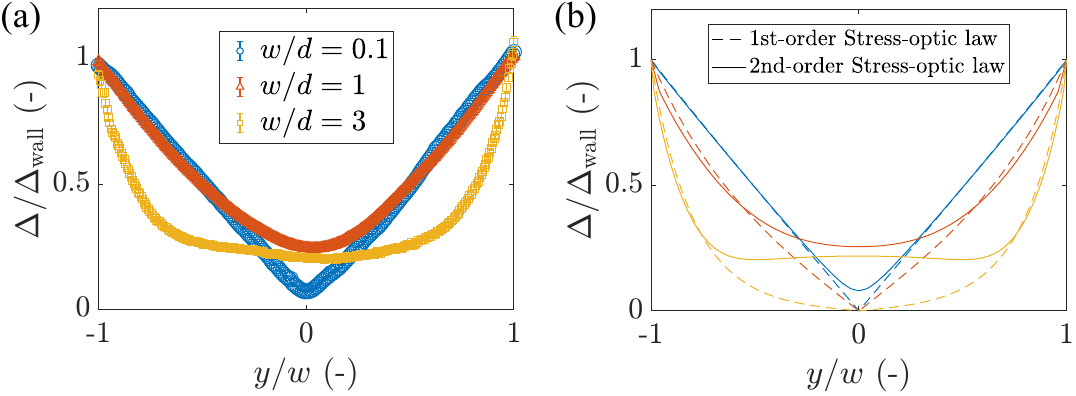}
\caption{\label{fig:retardation_comparisson} Comparison between (a) experimental and (b) theoretical retardation profiles for the 0.5 wt\% CNC suspension and different aspect ratios $w/d$. The retardation distributions $\Delta(y)$ are normalized by the retardation at the wall $\Delta_{\rm wall}$. The dashed and solid lines in (b) are calculated using 1st- and 2nd-order SOLs, respectively. For $w/d = 0.1$, the flow rate is $Q=20$ ml/min and we use $C_1 = 1.51\times 10^{-5}$ 1/Pa and $C_2 = 4.19\times10^{-4}$ 1/Pa$^2$ to calculate the theoretical retardation profile. For the case of $w/d = 1$, the flow rate is $Q=70$ ml/min and we use $C_1 = 1.52\times 10^{-5}$ 1/Pa and $C_2 = 1.11\times10^{-5}$ 1/Pa$^2$ to calculate the theoretical retardation profile. For the case of $w/d = 3$, the flow rate is $Q=50$ ml/min and we use $C_1 = 1.35\times 10^{-5}$ 1/Pa and $C_2 = 2.8\times10^{-5}$ 1/Pa$^2$ to calculate the theoretical retardation profile.}
\end{figure*}

\subsection{Retardation distributions}

The retardation fields were systematically measured using three rectangular channels and four CNC suspensions with different concentrations. We collected more than 100 data.
Figure \ref{fig:retardation_line} shows the line profiles of the measured retardation at each flow rate $Q$.
In all cases, the highest retardation values appear at the channel wall ($y = \pm w$) and the lowest values occur at the channel center ($y = 0$).
The retardation at each $y$-position increases with the flow rate.
At the wall ($y = \pm y$), the retardation for the same flow rate increases with the concentration $c$ of CNC.
This means that the concentration of CNC is related to the degree of birefringence, i.e., the stress-optic coefficient $C_1$ of Eq. (\ref{eq:Delta_wall}), that occurs at a given flow rate.
The spatial distributions of the retardation vary with the aspect ratio of the channel.
The rectangular channel with an aspect ratio of 0.1 exhibits a V-shaped retardation distribution with respect to $y/w$ because of the quasi-two-dimensional flow field.
The retardation distribution deviates from the V-shape as the aspect ratio increases.
The rectangular channel with an aspect ratio of 3 displays a plateau around $y = 0$ when the concentration $c$ or the flow rate is low.
For example, in Fig. \ref{fig:retardation_line}(c), the minimum retardation is approximately 1 nm.
This value is comparable to the retardation at rest, as shown in Fig.~\ref{Fig:retardation_at_rest}.
Therefore, the retardation may actually be lower than this value, but the camera noise prevents us from obtaining accurate measurements.
These results indicate that the use of CNC suspensions enables the measurement of flow birefringence distributions that reflect the aspect ratio $w/d$, i.e., the channel geometry, even at low CNC concentrations.

Equation (\ref{eq:Delta_center}) suggests that the retardation at the center of the channel, $\Delta_{\rm center}$, should be zero if there is no effect of the stress component along the camera's optical axis on the flow birefringence, i.e., $C_2 = 0$.
However, when the flow rate is sufficiently high, $\Delta_{\rm center}$ clearly shows a finite value and increases with the flow rate $Q$, even for the quasi-two-dimensional channel ($w/d = 0.1$) and at low CNC concentrations.
This result indicates that the term relating to the stress component along the camera's optical axis in the SOL must be considered in the flow birefringence, except when the flow rate is sufficiently low and when the channel flow is quasi-two-dimensional.

\subsection{Effect of stress component along the camera's optical axis on flow birefringence}

To quantitatively assess the effect of the stress component along the camera's optical axis on flow birefringence, the retardation values at the wall $\Delta_{\rm wall}$ and center $\Delta_{\rm center}$ of the channel are plotted against the flow rate $Q$ in Fig. \ref{fig:retardation_wall_center}.
The left-hand side of Fig. \ref{fig:retardation_wall_center} shows that the retardation at the wall increases in proportion to the 1st power of $Q$.
This result agrees with the expectation of both the 1st- and 2nd-order SOLs described in Eq. (\ref{eq:Delta_wall}).
Although a previous study suggested that retardation will reach a saturation level because of the saturated alignment of CNC \citep{detert2023,calabrese2023a} at sufficiently high flow rates, the relationship between retardation and flow rate remains linear within the range considered in this study.
The right-hand side of Fig. \ref{fig:retardation_wall_center} shows that the retardation at the center increases in proportion to the 2nd power of $Q$ if $Q$ is sufficiently large ($Q \geq 10$ ml/min in our study).
This result does not agree with the 1st-order SOL in Eqs. (\ref{eq:local retardation1}) and (\ref{eq:local retardation2}), but does agree with the expectation of the 2nd-order SOL in Eqs. (\ref{eq:v1}) and (\ref{eq:v2}), for which the stress component along the camera's optical axis, i.e., $C_2$, is nonzero.
These results indicate that the flow birefringence of CNC suspensions in a three-dimensional flow field can be better understood using the 2nd-order SOL, which considers the effect of the stress component along the camera's optical axis, rather than the 1st-order SOL that has conventionally been used.

As the lowest measurable retardation of the polarization camera used in this study is around $1.27$ nm, retardation values of less than around $1$ nm can be considered as noise.
For example, almost all $\Delta_{\rm center}$ for $w/d = 3$ are at the noise level because the stress along the $z$-axis and its integrating length are small for this case.
This region can be further verified using a long channel with a larger cross-section to gain a longer integrating length.
Additionally, $\Delta_{\rm center}$ is at the noise level for $w/d = 0.1$ with the 1.0 wt\% CNC suspension because of the low flow rate, $Q \leq 9$ ml/min.

To illustrate the practicality of the 2nd-order SOL, we now reexamine the retardation distribution.
Figure \ref{fig:retardation_comparisson} compares the experimental and theoretical retardation line profiles for all channel aspect ratios.
The retardation profiles are normalized by the retardation at the wall, $\Delta_{\rm wall}$.
In calculating the line profiles of the retardation using the 2nd-order SOL, $C_1$ and $C_2$ match the retardation at the wall and center of the channel with the respective experimental values.
The values of $C_2$ and $C_1$ for all cases are described in \ref{sec:stress-optic coefficient}.
Figure \ref{fig:retardation_comparisson} clearly shows that, for any aspect ratio, the retardation profiles calculated using the 2nd-order SOL, i.e., $C_2 \neq 0$, are in better agreement with the experimental results than those calculated using the 1st-order SOL, i.e., $C_2 = 0$.

\section{Conclusion}\label{sec13}

This study has attempted to clarify the effect of the stress component along the camera's optical axis on the flow birefringence of CNC suspensions by revisiting the SOL described by \citet{doyle1982nonlinearity}.
This effect has often been neglected in conventional 1st-order SOLs.
The 2nd-order SOL includes the stress component along the camera's optical axis.
In this study, systematic experiments were carried out to measure the flow birefringence of CNC suspensions in steady laminar flow through rectangular channels, and more than 100 datasets were assessed.
The retardation distributions were measured for various CNC suspensions (with 0.1, 0.3, 0.5, and 1.0 wt\% CNC concentrations) flowing through rectangular channels with aspect ratios of 0.1, 1, and 3.
The measured retardation values were compared with those predicted theoretically by the 1st- and 2nd-order SOLs.

The results show that the retardation distribution varies with the channel aspect ratio.
The 1st-order SOL holds in flow fields where the stress component along the camera's optical axis ($z$-axis) is negligible, such as in a quasi-two-dimensional flow channel with a small aspect ratio and low flow rate.
However, in cases where the stress component along the camera's optical axis cannot be ignored, such as in the case of a high flow rate, the 1st-order SOL cannot be applied, even for quasi-two-dimensional channel flows.
In such cases, the 2nd-order SOL predicts the retardation distribution with reasonable accuracy based on experimental data.
Focusing on the retardation at the channel center, where the effect of the stress component along the camera's optical axis is most pronounced, the 2nd-order SOL predicts the retardation to be proportional to the 2nd power of the flow rate $Q$, in agreement with the experimental retardation results.
To conclude, when measuring flow birefringence in the three-dimensional flow of CNC suspensions, the stress component along the camera's optical axis must be considered.
The results of this study will be of great importance in studies of flow birefringence in three-dimensional channels for CNCs and other materials.

\appendix
\def\thesection{Appendix \Alph{section}}

\section{Non-Newtonian effect of 1.0 wt\% CNC Suspension}\label{sec:non-Newtonian effect}

\begin{figure}[t]
\centering
{\includegraphics[width=1\columnwidth]{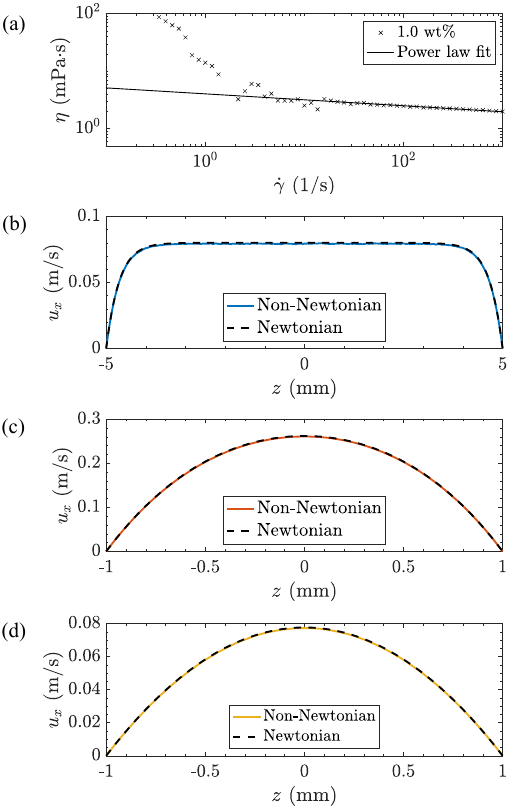}}
\caption{\label{fig:PowerLaw} (a) Viscosity of 1.0 wt\% CNC suspension. The solid line shows the power law model of $\eta = K \dot \gamma^{n-1}$ with $n \simeq 0.9$ and $K \simeq 4.0$. (b,c,a) are the velocity profiles along the $z$-axis at $y = 0$ when $Q=30$ ml/min for aspect ratios $w/d$ of 0.1, 1, and 3, respectively. The dashed lines show the theoretical velocity profile calculated by Eq. (\ref{eq:u}) using a constant viscosity of 1.58 $\rm mPa \cdot s$. The solid lines show the numerically calculated velocity profile of the power law fluid.}
\end{figure}

We now verify the non-Newtonian effect of the 1.0 wt\% CNC suspension.
The slight shear-thinning viscosity of the suspension was fitted by a power law model described by the following equation:
\begin{equation}
\eta = K \dot \gamma^{n-1},
\end{equation}
where $K$ is the flow consistency index and $n$ is the power exponent.
Figure \ref{fig:PowerLaw}(a) shows the viscosity of the 1.0 wt\% CNC suspension and the fitted power law model using $K \simeq 4.0$ and $n \simeq 0.9$.
The velocity profiles of the 1.0 wt\% CNC suspension flowing in rectangular channels with different aspect ratios were numerically calculated using commercial software (COMSOL Multiphysics) with the power law model and these constants.

Figures \ref{fig:PowerLaw}(b,c,d) show the measured velocity profiles (solid lines) for aspect ratios of 0.1, 1, and 3, respectively, and the theoretical distribution (dashed lines) calculated by Eq. (\ref{eq:u}) using a constant viscosity of 1.58 $\rm mPa \cdot s$.
The difference between the velocity distributions of non-Newtonian and Newtonian fluids is particularly small because the power exponent of 0.9 is close to 1.
Therefore, the non-Newtonian effect of the 1.0 wt\% CNC suspension is negligible.

\section{Stress-optic coefficients}\label{sec:stress-optic coefficient}

\begin{figure*}[t]
\centering
{\includegraphics[width=0.8\textwidth]{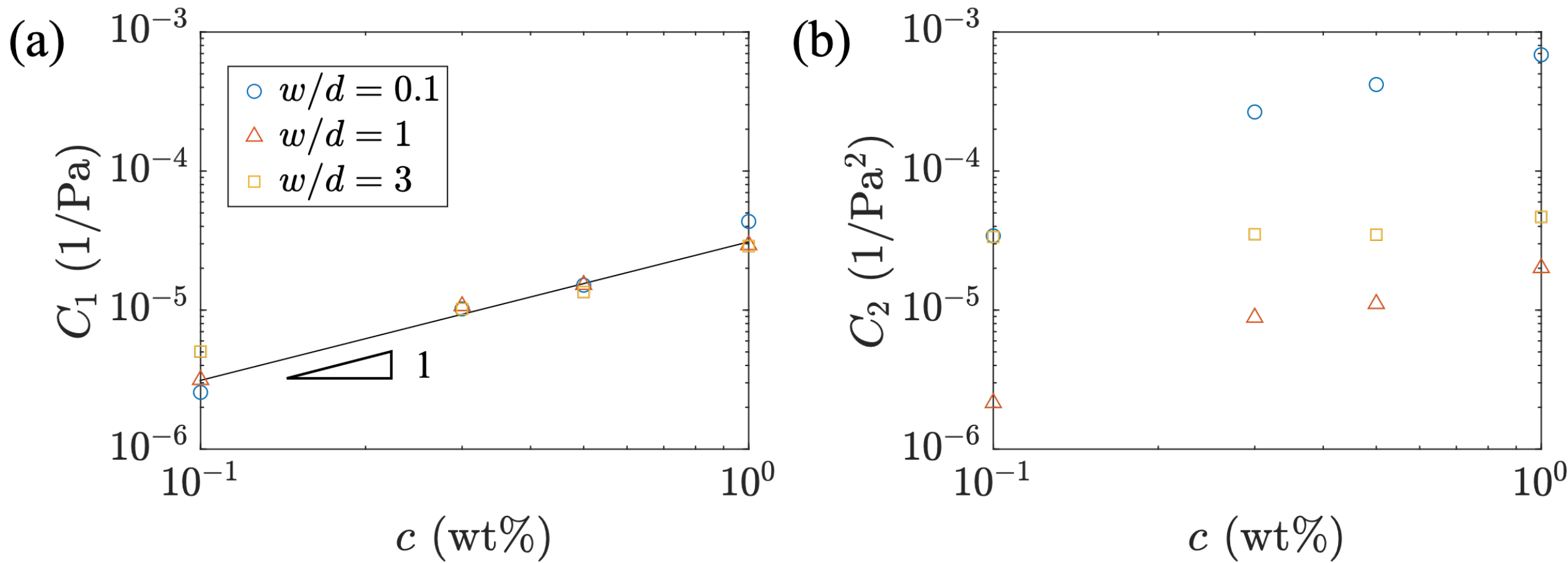}}
\caption{\label{fig:C1_C2} (a) $C_1$ and (b) $C_2$ adjusted to minimize the difference between the calculated theoretical retardation distribution and the experimental values at the wall and center of the channel, denoted as $\Delta_{\rm wall}$ and $\Delta_{\rm center}$, respectively. The solid line in (a) shows the linear relationship between $C_1$ and $c$, $C_1 \propto c^1$.}
\end{figure*}

Figure \ref{fig:C1_C2} shows the adjusted $C_1$ and $C_2$ such that the retardation at the wall and the center of the channel, $\Delta_{\rm wall}$ and $\Delta_{\rm center}$, are closest to the experimental value when calculating the theoretical retardation distribution.
$C_1$ is linearly proportional to the CNC concentration $c$ and is nearly independent of the aspect ratio $w/d$. 
The value of $C_2$ also increases with the CNC concentration, but varies with the aspect ratio.
This means that the magnitude of the effect of the stress component along the camera's optical axis on flow birefringence changes with the flow condition, because $C_2$ is the contribution level of the stress component along the camera's optical axis in the SOL.
Additionally, the procedure performed in this study provides a method for determining $C_1$ and $C_2$ in birefringent fluids.

\section*{Statements and Declarations}

\section*{Acknowledgments}

We thank Dr. T. Shikata, Dr. T. Rösgen, and Dr. C. Lane for their valuable discussions and suggestions. 

\section*{Funding}

This work was funded by Japan Society for the Promotion of Science KAKENHI (Grant Numbers: JP19K23483, JP20K14646, JP20H00222, JP20H00223, JP22KJ1239), Japan Science and Technology Agency PRESTO (Grant Number: JPMJPR21O5), and Japan Agency for Medical Research and Development (Grant Number: JP22he0422016).

\section*{Author contributions}

Y.T. and M.M. conceptualized and supervised the study.
K.N., Y.Y., W.W., and M.M. developed the methodology.
K.N., Y.Y., and W.W. conducted the formal analysis and investigation.
K.N., Y.Y., and W.W. prepared the original draft manuscript.
All authors reviewed and edited the manuscript.

\section*{Competing interest}

The authors have no conflicts of interest to declare that are relevant to the content of this article.

\section*{Data availability}

The data that support the findings of this study are available from the corresponding author upon reasonable request.

\bibliographystyle{spbasic} 
\bibliography{Ref}

\end{document}